%% file: paper.tex
\documentclass[conference]{IEEEtran}
\usepackage{times}
\usepackage[style=base]{caption} 
\usepackage{epsfig}
\usepackage{mathtools}
\usepackage{listings}
\usepackage{algorithm}
\usepackage{algpseudocode}
\usepackage{amsthm}
\usepackage{amssymb,amsmath}
\usepackage{ifthen}
\usepackage{color}
\usepackage[bookmarks=false]{hyperref}
\usepackage{arydshln}
\usepackage{scrextend}
\usepackage[inline]{enumitem}
\addtokomafont{labelinglabel}{\sffamily}

\theoremstyle{definition}

\newboolean{showcomments}
\setboolean{showcomments}{true}
\ifthenelse{\boolean{showcomments}}
{ \newcommand{\mynote}[3]{
    \fbox{\bfseries\sffamily\scriptsize#1}
    {\small$\blacktriangleright$\textsf{\emph{\color{#3}{#2}}}$\blacktriangleleft$}}}
{ \newcommand{\mynote}[3]{}}

\usepackage{cite}

\ifCLASSINFOpdf
\else
\fi
\begin{document}
%
\title{Alpha Entanglement Codes: Practical Erasure Codes to Archive Data in Unreliable Environments}



\author{\IEEEauthorblockN{Vero Estrada-Gali\~{n}anes\IEEEauthorrefmark{1},
Ethan Miller\IEEEauthorrefmark{2},
Pascal Felber\IEEEauthorrefmark{1} and 
Jehan-Fran\c{c}ois P\^aris\IEEEauthorrefmark{3}}
\IEEEauthorblockA{\IEEEauthorrefmark{1}University of Neuch\^atel,
2000 Neuch\^atel, Switzerland}
\IEEEauthorblockA{\IEEEauthorrefmark{2}University of California, Santa Cruz, CA 95064, USA}
\IEEEauthorblockA{\IEEEauthorrefmark{3}University of Houston, Houston, TX 77204-3010, USA}
}


\maketitle


\begin{abstract}
	\input{abstract}
\end{abstract}

%
\IEEEpeerreviewmaketitle

\input{introduction}
\input{background}
\input{alpha}
\input{system}

\input{evaluation}
\input{related}
\input{conclusion}

\section*{Acknowledgement}
The authors would like to thank the shepherd Yair Amir and the anonymous reviewers for their fruitful comments. 
V.E.G. thanks Jay Lofstead, Ahmed Amer, Kragen Sitaker, and Hugues Mercier for stimulating discussions, and to Luiz Barroso for answering questions regarding availability in global systems. 
This work was partially supported by SNSF Doc.Mobility 162014.



%
%
%

\bibliographystyle{IEEEtran}
\bibliography{./IEEEfull}

\end{document}

%% file: abstract.tex
Data centres that use consumer-grade disks drives and distributed peer-to-peer systems are unreliable environments to archive data without enough redundancy. 
Most redundancy schemes are not completely effective for providing high availability, durability and integrity in the long-term. 
We propose alpha entanglement codes, a mechanism that creates a virtual layer of highly interconnected storage devices to propagate redundant information across a large scale storage system. 
Our motivation is to design flexible and practical erasure codes with high fault-tolerance to improve data durability and availability even in catastrophic scenarios. 
By ``flexible and practical", we mean code settings that can be adapted to future requirements and practical implementations with reasonable trade-offs between security, resource usage and performance. 
The codes have three parameters. 
Alpha increases storage overhead linearly but increases the possible paths to recover data exponentially. 
Two other parameters increase fault-tolerance even further without the need of additional storage.
As a result, an entangled storage system can provide high availability, durability and offer additional integrity: it is more difficult to modify data undetectably.
We evaluate how several redundancy schemes perform in unreliable environments and show that alpha entanglement codes are flexible and practical codes. 
Remarkably, they excel at code locality, hence, they reduce repair costs and become less dependent on storage locations with poor availability. 
Our solution outperforms Reed-Solomon codes in many disaster recovery scenarios.  

%% file: introduction.tex

\section{Introduction}
Redundancy schemes play a key role in the security and performance of storage systems.
Many cloud-based centralised solutions use replication or RAID-like techniques~\cite{patterson1988case}, the latter often built with erasure codes~\cite{plank1997tutorial} like Reed-Solomon (RS) codes. 
In decentralised systems, replication is widely used due to the higher bandwidth requirements of erasure coding, particularly to rebuild single-failures. 
Choosing a redundancy scheme, for the most part, involves finding a good compromise between reliability, performance, and costs of additional storage, repair and maintenance of redundancy. 
The estimation of these variables is a subtle task, especially for archival systems that are expected to maintain data accessibility and integrity for the long term in unreliable environments. 

Some of the most remarkable works in the last 20 years of research propose scalable redundancy mechanisms to address availability and/or correlated failures: The Eternity Service (1996)~\cite{anderson1996eternity}, PASIS (2000)~\cite{wylie2000survivable}, FARSITE (2002)\cite{adya2002farsite}, Glacier (2005)~\cite{haeberlen2005glacier} and Carbonite (2006)~\cite{chun2006efficient} and recently RESAR (2016)~\cite{schwarz16-mascots}. 
In the same period of time, the storage scale has been changing abruptly and such change suggest us to rethink the schemes used to store data redundantly. 

The industry sector puts strong emphasis on storage costs, which means that their service usually rely on hybrid redundancy schemes. 
For instance,  Google uses multi-cell replication to survive data centre failures, and RS encoding for local failures~\cite{ford2010availability}.
Facebook uses geo-replicated XOR-based codes to provide fault-tolerance at the data centre level, and RS encoding within a single data centre~\cite{muralidhar2014f4}. 
RS codes are optimal storage codes, however, the overall storage overhead of the redundancy scheme is significant when used in combination with replication.

RS codes are designed to protect data against massive correlated failures, but, in practice, the parameter values are restricted to a small range due to poor performance. 
A system that uses \emph{RS(k,m)} codes tolerates $m$ failures, but requires $k$ I/O accesses and $k \cdot B$ bandwidth to repair a single failure of $B$ bytes.
Some solutions to reduce the repair overhead were proposed \cite{papailiopoulos2014locally, khan2012rethinking, huang2012erasure, sathiamoorthy2013xoring}.
Real-world systems advocate small values: RS(6,3) in Google's single-cell scheme ~\cite{ford2010availability}, RS(10,4) in Facebook's f4 system~\cite{muralidhar2014f4} and $k+m\leqslant20$ in Microsoft Azure ~\cite{huang2012erasure}.
With settings like those mentioned, systems can keep data safe only with the existence of continuous monitoring and repairing operations that increase storage cost.
Another limitation is that reliability requirements may change over time but RS parameters are not dynamic: data must be encoded again to change $k$ and $m$.  

In this work, we design a robust and flexible redundancy scheme that uses storage and bandwidth resources efficiently, accommodate future reliability needs and its implementation does not require ``magic numbers". 
We present alpha entanglement codes, AE($\alpha,s,p$), a family of erasure codes that is fundamentally centered around the new concept of redundancy propagation. 
Thus, an entangled storage system would provide permanent storage with high availability, durability and integrity despite external and internal threats. 
This work generalizes our earlier entanglement algorithms ~\cite{galinanes2013helical,estradasimple}, including them as special cases, defines redundancy propagation, discusses centralised and decentralised entangled storage systems, and improves evaluations done in other works~\cite{galinanes2015ensuring}. 

AE codes create redundancy by tangling (mixing) new data blocks with old ones, building entangled data chains that are woven into a growing mesh of interdependent content.
The robust connectivity provides high fault-tolerance.
The only assumption is that data are stored permanently, deletions are only possible at the beginning of the mesh.
The propagation is controlled by the code parameters:  
\begin{enumerate*}[label=(\roman*)]
\item $\alpha$ determines the local connectivity,  
\item $s$ and $p$ determine the global connectivity of data blocks in the grid. 
\end{enumerate*}
Single-entanglements ($\alpha = 1$) compute 1 parity per data block, double-entanglements ($\alpha = 2$) compute 2 parities,  triple-entanglements ($\alpha = 3$) compute 3 parities and so on. 
Although the storage overhead increases linearly with the number of parities per data block, the number of possible \emph{data recovery paths} grows exponentially. 
Due to the redundancy propagation, AE codes reveal a combination of features previously unavailable for storage systems. 

Alpha entanglements are irregular codes. 
In simple words, the fault tolerance of irregular codes~\cite{greenan08-dsn} goes beyond the $m$ failures tolerated in $(k,m)$-codes but the extra tolerated failures are not arbitrary. 
The usual metrics to measure irregular fault tolerance are based in the distribution of the failure patterns that the code cannot tolerate. 
We use a variation of previously proposed metrics~\cite{wylie2007determining,greenan08-dsn} to show the benefits of redundancy propagation. 
Larger values in $\alpha$, $s$, and/or $p$ increase the size of minimal erasure patterns, $|ME(x)|$, that cause the irrecoverable loss of $x$ data blocks. 
For example, we will show that a pattern that cause the loss of two data blocks for code setting AE(3,1,4) is $|ME(2)|=8$ and it becomes considerable larger, $|ME(2)|=14$, for code setting AE(3,4,4).
A remarkable aspect of redundancy propagation is that tuning the parameters ($s$,$p$) does not modify the storage overhead and that none of the three parameters can change the cost of a single failure, which is always repaired by XORing two blocks. 
Another characteristic is that alpha entanglements permit changes in the parameters without the need to encode the content again. 
This property opens the possibility of a dynamic fault-tolerance, which is an interesting feature for long-term storage systems. 
As far as we know, these characteristics are not found in state-of-the-art codes.

The next section provides the background and the motivation for this work. 
We present alpha entanglement codes in Section~\ref{alpha} and provide a guideline of how they can be adapted to diverse applications and storage architectures in Section~\ref{system}. 
The results of our evaluation are presented in Section~\ref{evaluation}. 
First, we investigate the impact of different code settings on fault tolerance and write performance.  
Second, we run simulations over millions of synthetically generated blocks to evaluate data loss, redundancy degradation and repair performance in catastrophic failures to compare alpha entanglement codes and $(k,m)$-codes. 
Finally, we show that alpha entanglement codes outperform Reed-Solomon codes when are used in unreliable environments with poor maintenance. 
Alpha entanglement codes excel at code locality, hence they reduce repair cost and are less dependent on storage locations with poor availability.  

%% file: background.tex

\section{Background and Motivation}
Mixing data is being actively investigated in network coding ~\cite{dimakis2006decentralized,gkantsidis2005network,zhang2006benefits} as well as in storage applications ~\cite{aspnes2007towards,tran2008friendstore,waldman2001tangler,stubblefield2001dagster}.
More specifically, document entanglement~\cite{waldman2001tangler,stubblefield2001dagster} was proposed for censorship resistant systems  but as pointed by others an efficient and practical method is not straightforward to devise~\cite{aspnes2007towards}.
We revisited the problem in the context of fault-tolerant systems and proposed helical entanglement codes (HEC)~\cite{galinanes2013helical}. 
The code computes the exclusive-or (XOR) of two blocks: a parity and a data block. 
Every new data block is entangled with three parities. 
The three outputs contribute to enlarge three specific strands (chains of interleaved data and parity blocks) determined by the algorithm.
The pattern is repetitive, i.e. the new parity block is used together with a newcomer block as input for the next XOR operation. 
Hence, the algorithm propagates redundant information in regular patterns into disks arrays. 
The code is also known as $p$-HEC because its $p$ double-helix strands that resemble a DNA topology. 
The code builds, in total, $2$ horizontal and $2\times p$ helical strands. 
In a short paper, we gave some insights into the complex calculation of reliability in entangled storage systems by analysing 5-HEC reliability with a hierarchical decomposition of serial and parallel subsystem ~\cite{galinanes2015ensuring}. 
Recently, we proposed a simpler algorithm, which merely uses one single strand, showed that entanglements can provide more fault tolerance than mirroring arrays ~\cite{estradasimple}.
The early evidence of high reliability in entanglement codes is promising. 
Hence, it has motivated our study to answer important questions about redundancy propagation and its impact on a storage system.

Alpha entanglement codes provide a general model and unifying theory for the design of practical entanglement codes.  
We define $\alpha$-entanglement families and evaluate them in detail. 
This work significantly extends previous studies on entanglement codes, which become particular entanglement cases of AE codes. 
The previous defined $p$-HEC method corresponds to the family of triple-entanglements with $s=2$, in short AE($3,2,p$). 
Our motivation is to design practical erasure codes for large scale environments instead of another code based on traditional erasure codes originally designed for data transmission over erasure channels. 
We believe that, as indicated by other authors~\cite{oggier2013coding}, the translation of communication channel into a storage device is a simplification that hides maintenance costs in long-term storage.
This awareness motivated this work that investigates further on entanglement models and their implementation to build an entangled storage system. 

%% file: alpha.tex

\section{Alpha Entanglement Codes}
\label{alpha}
\textit{$\alpha$-entanglements($s,p$)} is a family of erasure codes to tangle data and redundant blocks with the goal of increasing the scope of redundant information by means of propagation. 
The encoder builds block chains, strands, that alternate data and redundant blocks.
The entanglement function computes the exclusive-or (XOR) of two consecutive blocks at the head of a strand and inserts the output adjacent to the last block. 
The strands are intertwined creating a mesh of entangled blocks.

\begin{figure}[!tb]
\centering
\includegraphics[width=0.85\columnwidth]{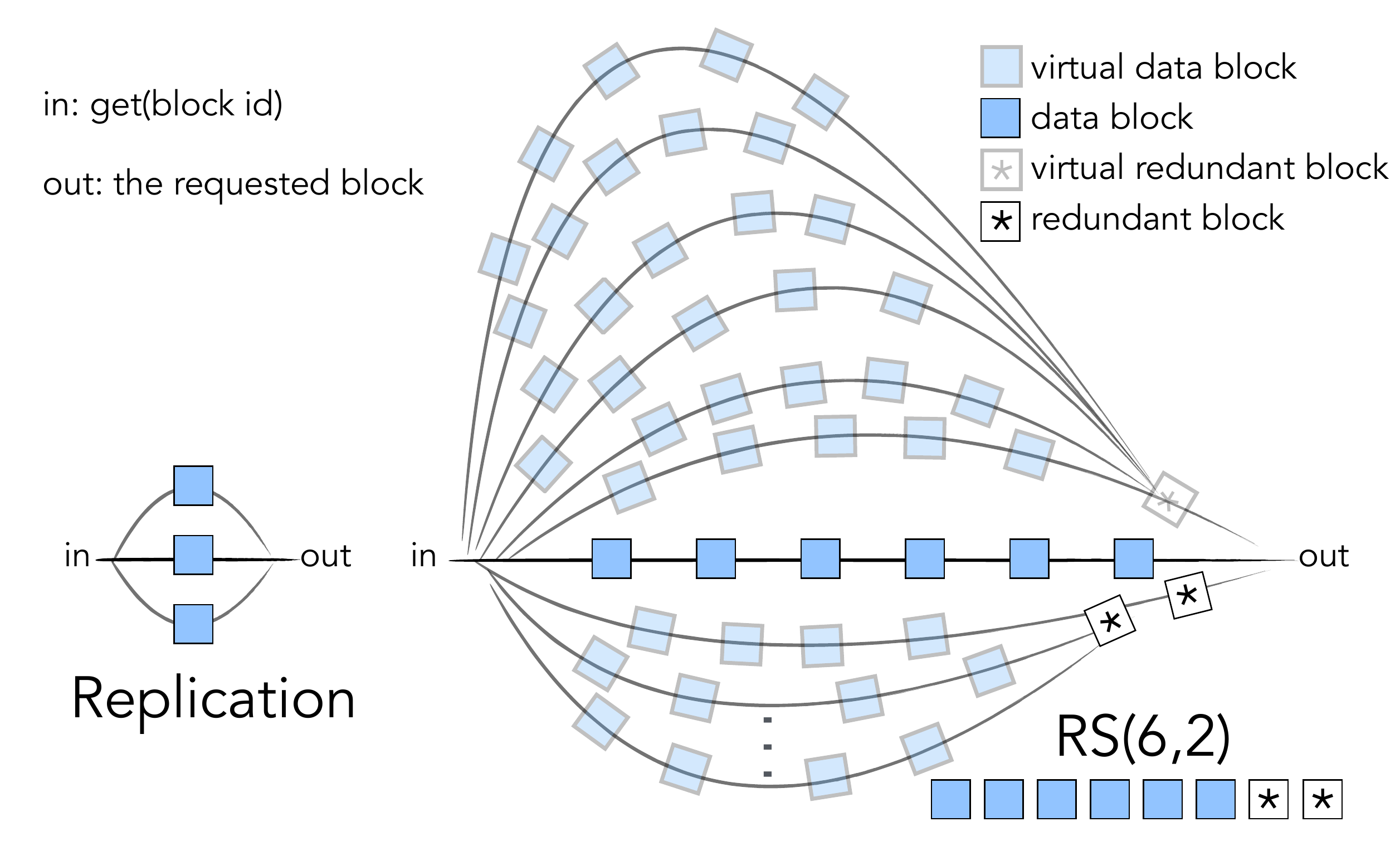}
\caption{Replication creates parallel paths, any of them can satisfy the request. 
RS codes create parallel and serial paths, e.g., RS(6,2) splits the source in $6$ blocks and expands it with $2$ redundant blocks. Any combination of $6$-out-of-$8$ blocks is a valid output. 
Parallel paths are constructed with virtual blocks.}
\label{fig:classic_redundancy}
\centering
\includegraphics[width=0.85\columnwidth]{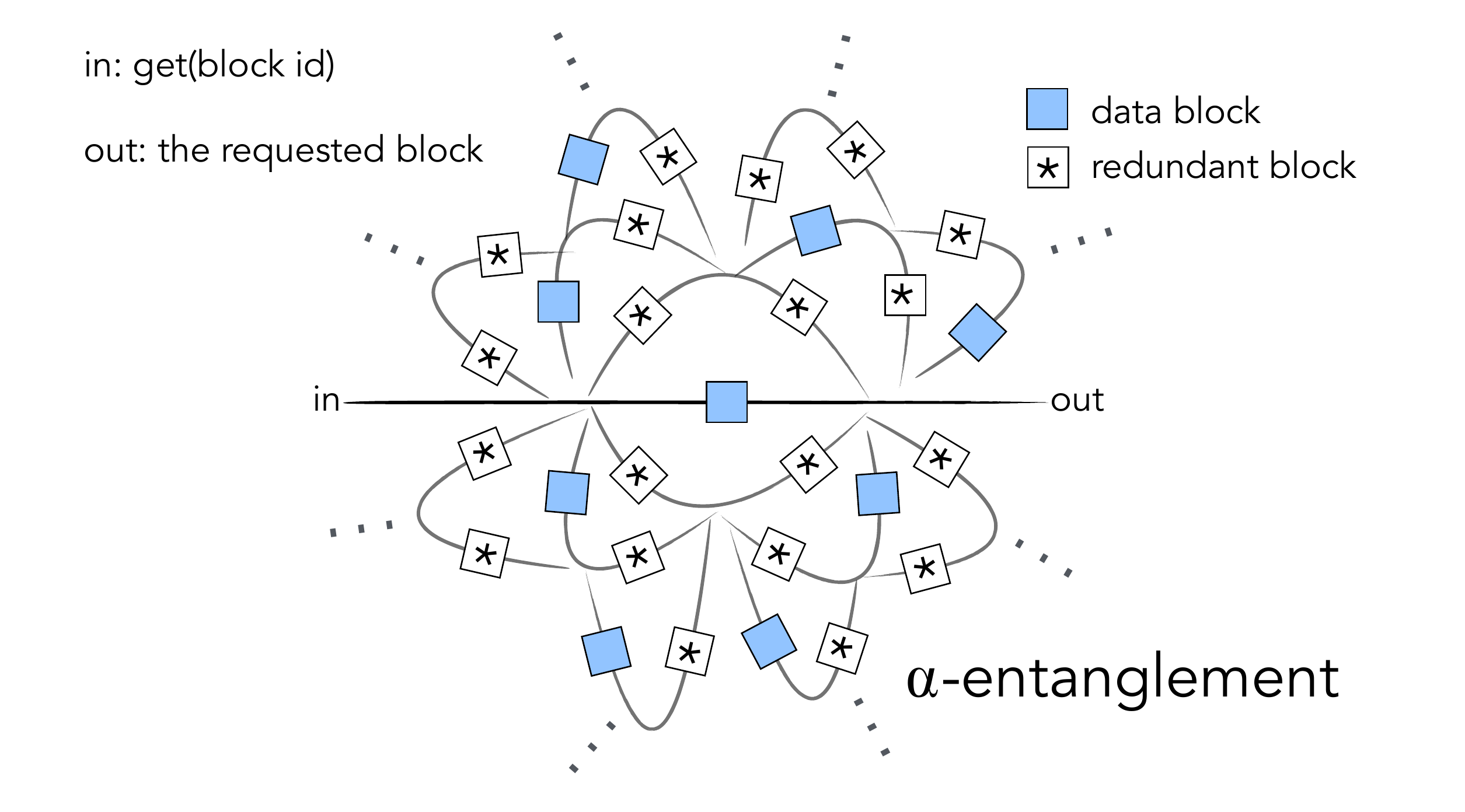}
\caption{AE codes create concentric/parallel paths for each data block.  Any path that connects in-out is valid to read the central block. Paths that are closer to the center have less elements in serial combinations while paths that are further away require more blocks but increase the chances of success.}  
\label{fig:novel_redundancy}
\end{figure}
\subsection{What Is Redundancy Propagation? Why We Should Care?}
Our notion of redundancy propagation challenges the traditional notion of redundancy. 
In engineering, the redundancy of a system is explained in terms of how the components are arranged in serial and parallel combinations. 
If elements are arranged in series, it means that the system needs all elements to satisfactorily perform the task.
Elements that are arranged in parallel increase the chances that the system will operate successfully. 
Indeed, an encoder algorithm creates paths that connect data blocks in serial and parallel arrangements. 
A decoder algorithm uses the paths to decode data.    

Fig.~\ref{fig:classic_redundancy} illustrates the classic methods to store data redundantly in a system. 
Replication improves reliability by creating $n$ parallel paths of single blocks.
RS codes and RAID-like codes improve performance by using serial paths of $k$ blocks. 
They also improve reliability by creating virtual parallel paths since any combination of $k$-out-of-$n$ blocks, with $n=k+m$, are useful to read the $k$ data blocks. 
The paths are virtual since blocks are not replicated, hence, this method is storage efficient.
Though, it is not optimal in terms of bandwidth and I/O operations, in particular for single failure repairs.  

Fig.~\ref{fig:novel_redundancy} illustrates the concept of redundancy propagation. 
Any arbitrary data block propagates redundant information to other blocks. 
The figure shows many valid paths to read the block located at the center of the image.  
The block is at the center of its loosely called \emph{propagation sphere}. 
The entanglement codes create, with a few simple rules, an arrangement of overlapping-spheres packing (each data block shown in the figure is at the core of its own propagation sphere, which are not shown for simplicity). 
In spite of the overlapping, each data block is uniquely identified by its position in a lattice, which determines unique combinations of two or more blocks that are useful to read the block.
The entanglement process creates concentric paths for each data block. 
The paths that are close to the center have less elements in serial combinations while the paths that are more distant require more blocks. 
The decoder uses the shortest available path to repair a missing data block. 
It rarely needs to use a long path since single failures are efficiently repaired using two blocks.
Since the decoder can repair multiple single failures in parallel, a long path affected by two or more failures may be repaired in a single round. 
Otherwise, the decoder tries again in the next round.
Redundancy propagation increases the probability of successfully recovering from catastrophic failures. 
It also makes difficult tampering with data because information is propagated to many elements in the system.

\subsection{How Do We Achieve Redundancy Propagation?}
\begin{figure}[t!]
\centering
\includegraphics[scale=0.28]{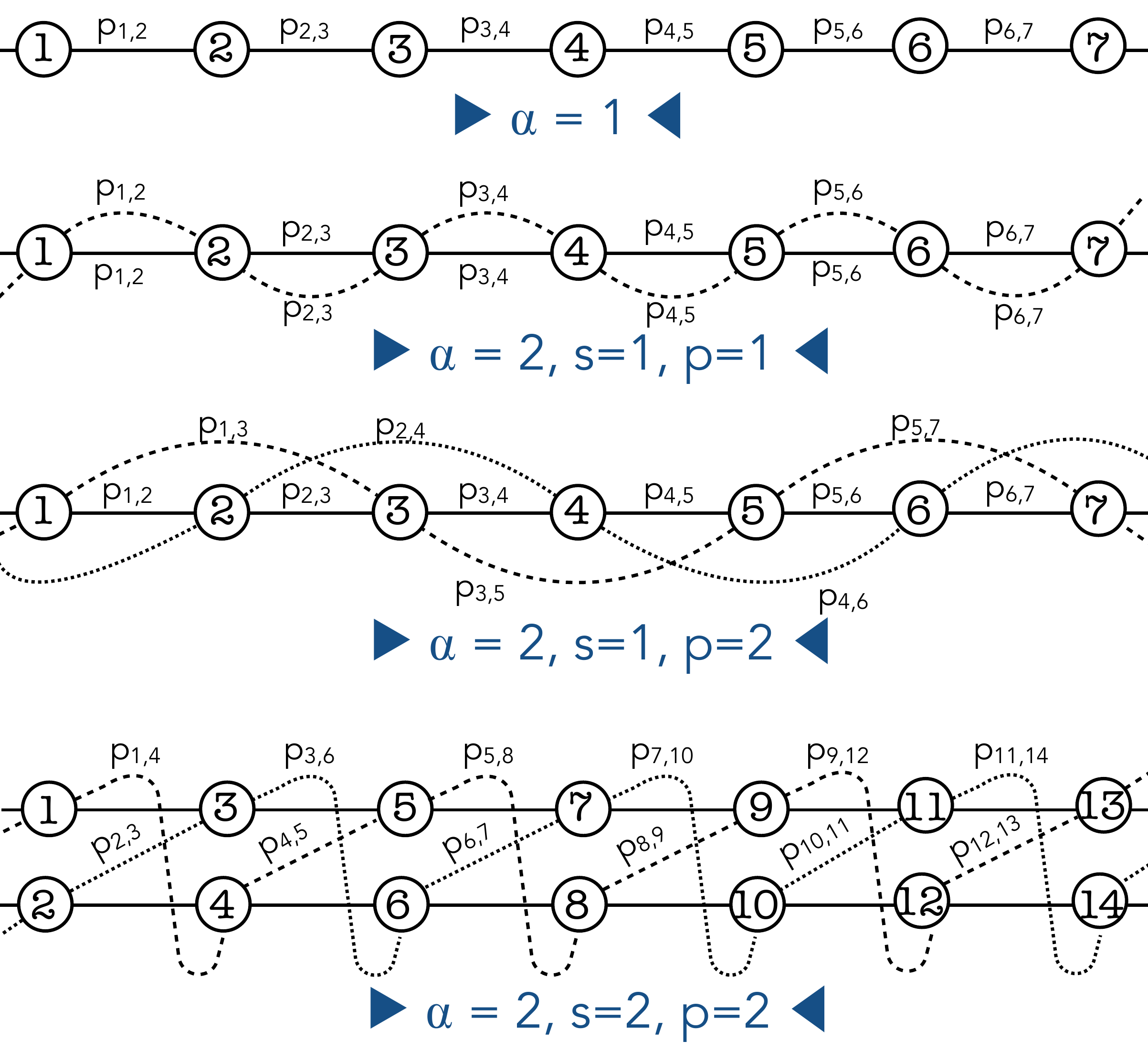}\\
\caption{\emph{Single- and double-entanglements.} 
A node, $d_i$, represents a data block and an edge, $p_{i,j}$, represents a parity block. Nodes are added to the graph in sequential order. 
They are uniquely identified by its position $i$, indicated inside the nodes.}
\label{fig:alfa_entanglements}
\end{figure}

AE codes are a family of codes composed by:
\begin{enumerate*}[label=(\roman*)] 
\item Single entanglements or \emph{1-entanglements}, $\alpha = 1$, built with a single horizontal strand. 
\item Double entanglements or \emph{2-entanglements(s,p)}, $\alpha = 2$, built with $s$ horizontal strands and one class of $p$ helical strands.
\item Triple entanglements or \emph{3-entanglements(s,p)}, $\alpha = 3$, built with $s$ horizontal strands and two classes of $p$ helical strands.
\item n-Tuple entanglements or \emph{n-entanglements(s,p)}, $\alpha = n$, built with $s$ horizontal strands and $n - 1$ classes of $p$ helical strands.
\end{enumerate*} 
Fig.~\ref{fig:alfa_entanglements} shows examples.  
A node, $d_i$, represents a data block and an edge, $p_{i,j}$, represents a parity block. 
The entanglement creates new parities and builds strands by adding elements sequentially. 
Strands are intertwined forming a graph.
A node is uniquely identified by its position $i$ in the graph. 
An edge $p_{i,j}$ connects nodes $d_{i}$ and $d_{j}$. 
It represents a parity block that is computed by XORing the last parity and data block of the same strand. 

\noindent
\textbf{Strands.} Each individual strand corresponds to a single entanglement. 
The complex combination of strands brings additional properties and increases fault tolerance. 
One of the emergent properties of $\alpha$-entanglements is that failure patterns that are not tolerated with single entanglements become innocuous in entanglements with $\alpha>1$ since data blocks are reconstructed through any of the $\alpha$ strands in which they participate. 
We define three classes of strands, the horizontal (H), the right-handed (RH) and the left- handed (LH) strands. 
In a 3D space, the helical strands revolve around a central horizontal axis and grow towards the right direction.
This work considers helical strands that connect horizontal strands with a diagonal of slope 1. 
If possible, there is a balanced quantity of RH and LH strands. 
In other words, 2-entanglements are composed with one class of helical strands (RH or LH) and 3-entanglements are composed with both classes of strands. 

A lattice is composed by s horizontal strands, and $\alpha -1 $ helical strand classes. 
Hence, the total number of strands is given by the formula $s+(\alpha - 1)\cdot p$.

\noindent
\textbf{Code Parameters.} 
Data propagation is tailored with three parameters: \textit{$\alpha$}, \textit{s}, \textit{p}.
The parameter $\alpha$ specifies the number of parities created per data block. 
It also describes the number of strands in which one data block participates.
Tuning $\alpha$ impacts on the resilience of the structure. 
More parities create more redundancy, on the other side, the storage overhead increases. 
The parameter $\alpha$ determines the code rate, which is computed with $\frac{1}{\alpha+1}$. 
Optionally, systems that only store parities have an improved rate of $\frac{1}{\alpha}$.
This work focuses on codes with $\alpha \in [1, 3]$ and gives some guidelines for large values of $\alpha$.
The parameter $p$ specifies the number of helical strands, and the parameter $s$ defines the number of horizontal strands.
Increasing $s$ and/or $p$ impacts positively on the structure's resilience without adding storage overhead. 
 
The parameter $\alpha$ and the lattice geometry put some constraints on the parameters.
1-entanglements are formed with only one chain of entanglements, therefore, $s = 1$ and $p = 0$. 
$\alpha$-entanglements with $\alpha>1$ are valid when $p \geq s$. 
An invalid setting, i.e. $p < s$, causes a deformed lattice.

\begin{table}[!tb]
\setlength{\tabcolsep}{2pt}
\small
\caption[Entanglement Rules - Input in Triple-Entanglements]{3-entanglement INPUT rules.  AE(3,5,5) example: on RH strand top node $d_{26}$ is tangled with $p_{25,26}$}
\centering
\begin{tabular}{r|c|c|c|c}
    & 
    \multicolumn{3}{c }{\bf $d_i$ is tangled with $p_{h,i}$, and $h$ index is} \\
    \bf $d_i$ position& 
    \multicolumn{1}{c|}{\bf H strand} &
    \multicolumn{1}{c|}{\bf RH strand} &
    \multicolumn{1}{c }{\bf LH strand} \\ \hline
    
    \bf top& 
    $i-s$ &    $i-s \cdot p+(s^2-1)$      & $i-(s-1)$ \\ \hline
    
    \bf central& 
    $i-s$          &   $i-(s+1)$  & $i-(s-1)$ \\ \hline
    
    \bf bottom & 
    $i-s$ &   $i-(s+1)$ & $i-s \cdot p+(s-1)^2$ \\ \hline
\end{tabular}

\label{table:CE_encoder_input}
\vspace{5pt}
\setlength{\tabcolsep}{2pt}
\small
\caption[Entanglement Rules - Output in Triple-Entanglements]{3-entanglement OUTPUT rules. AE(3,5,5) example: on RH strand top node $d_{26}$ entanglement creates $p_{26,32}$}
\centering
\begin{tabular}{r|c|c|c|c}
    & 
    \multicolumn{3}{c }{\bf $d_i$ entanglement creates $p_{i,j}$, and $j$ index is} \\
    \bf $d_i$ position& 
    \multicolumn{1}{c|}{\bf H strand} &
    \multicolumn{1}{c|}{\bf RH strand} &
    \multicolumn{1}{c }{\bf LH strand} \\ \hline
    
    \bf top& 
    $i+s$ &    $i+s+1$      & $i+s \cdot p-(s-1)^2$ \\ \hline
    
    \bf central& 
    $i+s$          &   $i+s+1$  & $i+s-1$ \\ \hline
    
    \bf bottom & 
    $i+s$ &   $i+s \cdot p-(s^2-1)$ & $i+s-1$ \\ \hline
\end{tabular}
\label{table:CE_encoder_output}
\end{table}
\begin{figure}[t!]
\centering
\includegraphics[width=\columnwidth]{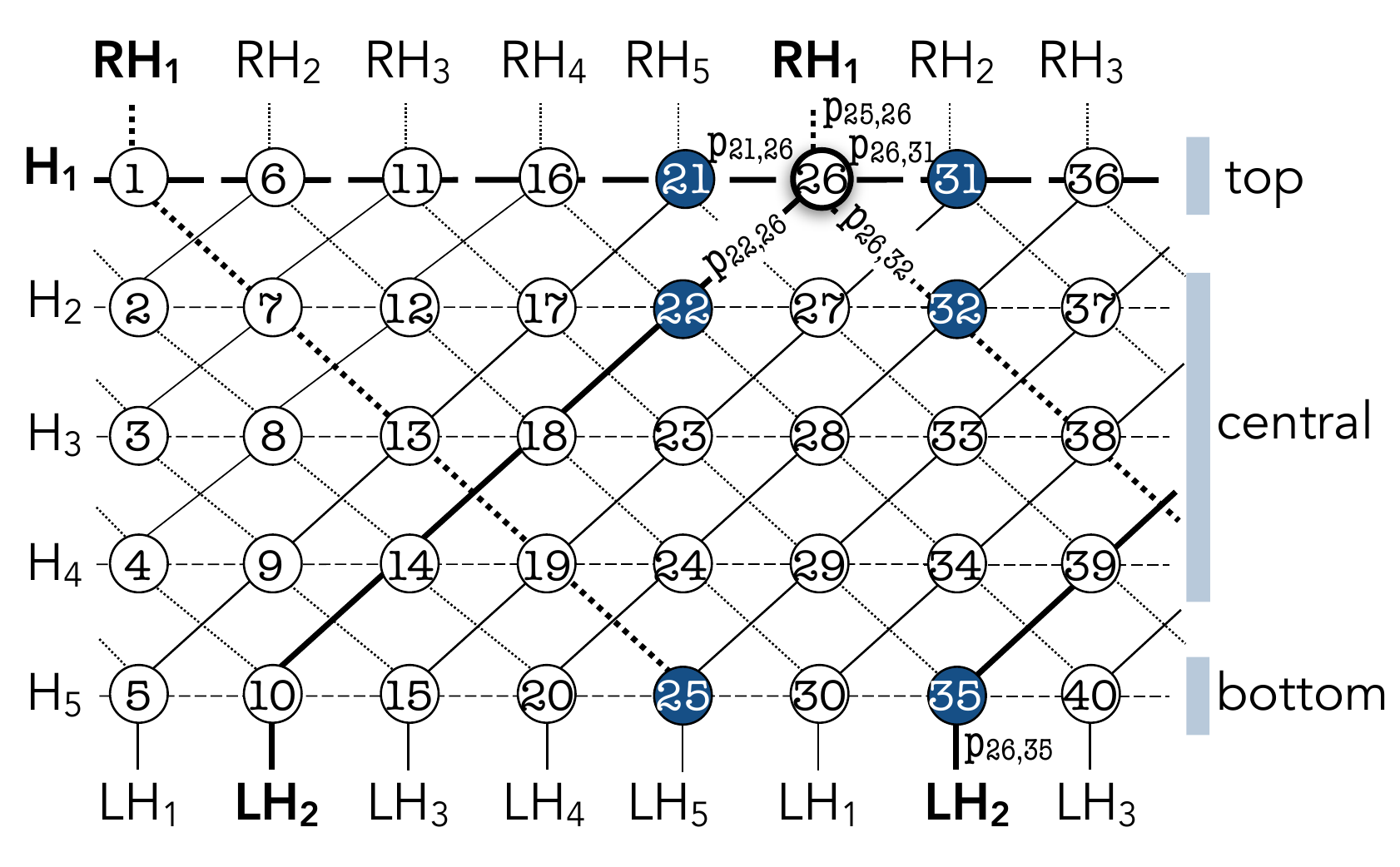}\\
\caption{AE(3,5,5) lattice with s=5 (rows) and p=5 (columns/diagonals). Any node $d_i$ participates in $\alpha=3$ strands, e.g. $d_{26}$ is a top node that belongs to $H_1$, $RH_1$ and $LH_2$ strands. Colored nodes are at one hop of node $d_{26}$.}
\label{fig:lattice}
\end{figure}
\noindent
\textbf{Code Specification.}
The encoder constructs a helical lattice using data and parity blocks with identical size. 
Data blocks are represented by nodes, and parities are represented by edges. 
Each node belongs to $\alpha$ strands, each edge belongs to only one strand. 
There are three categories: top, central and bottom nodes that determine the encoder rule. 
The encoder processes data blocks iteratively and keeps a counter $c$ that indicates the last processed block. 
First, a new data block is assigned an index $i=c+1$ that indicates its lattice's position. 
We refer to this data block as $d_i$. 
Second, the node category is determined by: \emph{top} if $i \equiv 1 \mod s$, \emph{bottom} if $i \equiv 0 \mod s$, and \emph{central} if $i > 1 \mod s$. 
Third, the encoder computes $\alpha$ parities for block $d_i$. 
Each new parity $p_{i,j}$ is computed by XORing the data block $d_i$ with an existing parity block $p_{h,i}$, whose index $h$ and $j$ are determined using rules tables.  
Fig.~\ref{fig:lattice} shows a lattice created with AE(3,5,5) using the rules given in Tables~\ref{table:CE_encoder_input} and~\ref{table:CE_encoder_output}.
The lattice is composed by 15 strands: 5 H strands $H_{1-5}$, 5 RH strands $RH_{1-5}$, and 5 LH strands $LH_{1-5}$. 
The decoder repairs a node using two adjacent edges that belong to the same strand, thus, there are $\alpha$ options. 
To illustrate, to repair $d_{26}$ from Fig.~\ref{fig:lattice} using H strand, the decoder computes the XOR($p_{21,26}$,$p_{26,31}$), whose indexes are determined with Tables~\ref{table:CE_encoder_input} and~\ref{table:CE_encoder_output} respectively.    
The decoder repairs an edge using any of the two incident nodes on the damaged edge and its corresponding adjacent edge, hence, there are always two options.
For instance, using the same figure, to repair $p_{21,26}$,  it computes the XOR($d_{21}$,$p_{16,21}$).

\noindent
\textbf{Implementation Details.}
The lattice acts as a virtual layer on top of the storage devices, or storage nodes in a p2p application. 
As with any other redundancy method, storage systems use mapping algorithms to store and locate encoded blocks according a placement policy and the available resources.  
The encoding and decoding implementation may use a client-based, middleware-based, or backend-based approach.    

\noindent
\textbf{Reducing Storage Overhead.}
The storage overhead does not grow gradually, i.e., increasing $\alpha$ in one unit means that the storage overhead increases in steps of 100\%.
We propose two strategies to enhance the code rate. 
A first option is to start with a low $\alpha$ and increase the value later as required.
A second option is to puncture the code. 
Puncturing is a standard technique use in coding theory in which, after encoding, some of the parities are not stored in the system. 
We will report our findings on this subject in the near future. 

\noindent
\textbf{Anti-tampering Property.}
The protection against data manipulation is an emergent property. 
To go undetected, an attacker should modify the $\alpha$ strands in which the targeted block participates by replacing all the parities computed from its position to the closest strand extremity. For example, to tamper $d_{26}$ in AE(3,5,5) shown in Fig.~\ref{fig:lattice}, the attacker needs to recompute $d_{26,31}$, $d_{31,36}$ and all the parities on the strand until the end of $H_1$ and do the same for $RH_1$ and $LH_2$. 

%% file: system.tex

\section{Entangled Storage System}
\label{system}
In this section, we explain how AE codes can be adapted to diverse applications without being restricted to a particular storage architecture. 
We describe two cases: 
\begin{enumerate*}[label=(\roman*)] 
\item a cooperative storage system built on top of a decentralised database, and 
\item a centralised storage architecture (disk arrays). 
\end{enumerate*}

\subsection{Use Case: A Geo-Replicated Backup}
A community creates a cooperative storage network to share storage and bandwidth resources. 
Users keep their own data in their local computers (nodes) and upload redundant information to geographically distributed nodes. %
The storage capacity used by each user has to be agreed at the beginning and renegotiated as needed.
The system is a two-tiered architecture that aims at supporting efficient data protection and integrity. 
The lower tier is composed of \emph{storage nodes} that share space to store parity blocks from other users. 
The upper tier consists of \emph{broker nodes} that encode and decode data.
In this example a single node has both roles a broker and storage node. 
The nodes are organized in a loosely connected cluster. 
In order to avoid confusion with the helical lattice graph, the text clarifies if ``node" refers to a graph vertex; otherwise we use the terms \emph{d-block} for a vertex and \emph{p-block} for an edge.
It is worth noticing that the final design requires consideration of various types of attacks such as free riding abuses that are beyond the scope of this work. 
In particular, methods to defend against free riding were proposed previously by other authors~\cite{Lillibridge:2003:CIB:1247340.1247343}.\\
\noindent
\textbf{Redundancy Scheme.}
Each user manages his own entanglement lattice via the broker, hence multiple lattices coexist in the system.
Moreover, the system could keep lattices with different settings. 
In the simplest scenario, the broker is a service running in the user's computer.
In more complex scenarios, the broker could be a super node that encodes data on behalf a group of users.
The broker recognizes files for backup. 
First, it prepares the file by splitting it into d-blocks. 
To entangle them, it needs to fetch some parities (p-blocks) from remote nodes. 
For performance, it can keep the parities from previous encodings in memory.
In that case, the memory footprint of the broker is linear in the number of distinct strands that create the entanglement lattice. 
For instance, AE(3,5,5) requires to keep in memory the last $p$-block of its 15 strands.
If the broker crashes, it only needs to retrieve the p-blocks from the remote nodes. 
Blocks are located by their key, which can be a value derived from the node id and the block position in the lattice (such as a hash of both values). 
Parities are mapped to the storage nodes using a deterministic or random placement algorithm.
In a failure-free environment, users can access their data directly from their local computers. 
In other words, decoding is not required.\\
\noindent
\textbf{Failure Mode.}
\begin{figure}[t!]
\centering
\includegraphics[width=\columnwidth]{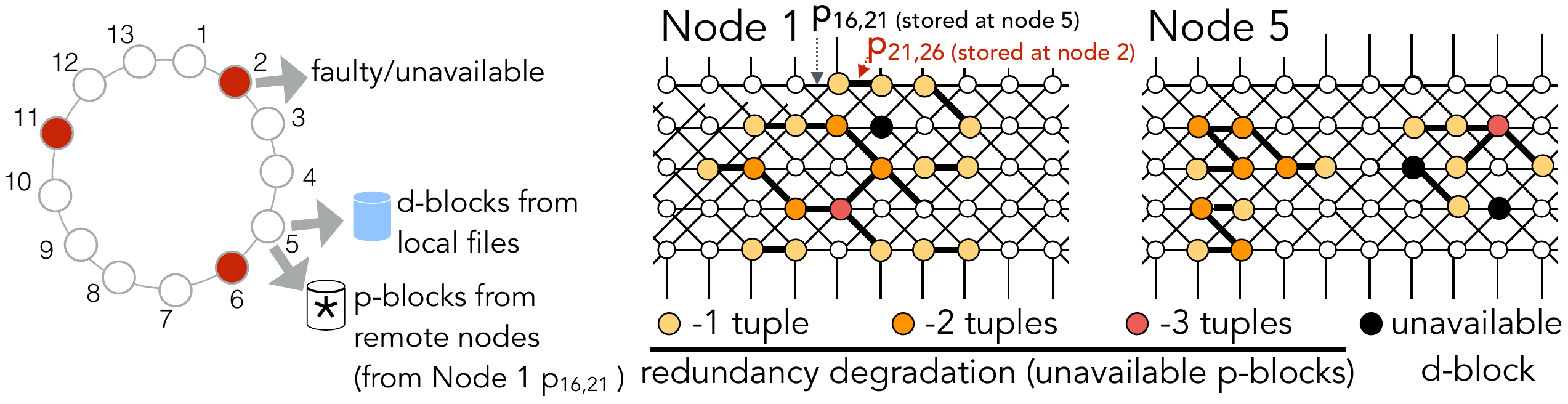}\\
\caption{\emph{Failure mode.} In a healthy AE(3,5,5) lattice each d-block has three pp-tuples stored at remote nodes. Unavailable nodes degrade redundancy. Each lattice is affected differently.}
\label{fig:failure_mode}
\end{figure}
Repairs occur when users do not have access to their local d-blocks or when their lattices deteriorate due to faulty or unavailable nodes. 
How node failures are detected and notified is not discussed here but there is plenty of literature on this topic. 
Fig.~\ref{fig:failure_mode} shows a scenario in which three nodes are unavailable and illustrates how the incident impacts on multiple elements in distinct lattices.
The size and characteristics of the incident dictate the patterns formed on the lattices. 
Parity block repair is automatically distributed, assuming that all users will be interested in the regeneration of their lattices to maintain the same level of redundancy for their data. 
If a node is not able to repair the lattice, other nodes can do repairs on their behalf as well. 

Repairing d-blocks requires complete pp-tuples (two p-blocks) and repairing p-blocks requires complete dp-tuples (one d-block and one p-block).  
Fig.~\ref{fig:failure_mode} indicates redundancy degradation. 
Some nodes have incomplete tuples. 
Each unavailable block generates a single-failure repair.
Table~\ref{tab:repair_lattice} enumerates the steps to regenerate the parities located in faulty nodes and illustrates with an example. 
Node 1 performs the steps 1-3 and 5 locally, and node 5 answers step 4. 

\begin{table}[t!]
\setlength{\tabcolsep}{2pt}
\small
\caption{Node 1 recomputes $p_{21,26}$ after failure.}
\centering
\begin{tabular}{lc}
\hline
\textbf{Steps} & \textbf{Example}     \\\hline
\textbf{1. Obtain dp-tuple id:}& $\{key_{21},key_{16,21}\}$,$\{key_{26},key_{26,31}\}$ \\\hdashline 
\textbf{2. Choose p-block id:} &$key_{16,21}$\\\hdashline
\textbf{3. Compute location key:} &$n_5$ \\\hdashline
\textbf{4. Get block:}& $p_{16,21}$\\\hdashline
\textbf{5. Repair block:} &$p_{21,26}$ \\
\hline
\end{tabular}
\label{tab:repair_lattice}
\end{table}

\subsection{Use Case: Disk Arrays}
Disk arrays can increase the performance of a system because individual requests are served faster when multiple disks collaborate to increase throughput. 
In addition, the possibility to answer multiple requests in parallel gives the chance to increase the system's I/Os rate. 
Some form of redundancy is required to compensate the increase of combined disk failure rate when writing to disk arrays. 
RAID organizations~\cite{patterson1988case} are a well-established solution for the industry and the home user to increase the performance and reliability of large arrays of inexpensive disks. 
Although there is still future for cheap spinning disks in data centres and home solutions, the market changed substantially in the last decades. 

Storage at exabyte scale brings difficult challenges:
\begin{enumerate*}[label=(\roman*)]
\item \emph{Rebuild time} may take hours or days. Systems that tolerate 2 failures dominate the market. More failures during rebuilds are a source for data loss. 
\item The \emph{disparity between technology development} means that disk capacity has been grown steadily while bandwidth lays behind.
\item \emph{Cloud-based storage has received more research attention than RAID solutions}, even though RAID is still in use for many reasons, e.g. privacy. 
\item \emph{Failure rate} is mostly related to system's size~\cite{schroeder2010large}. 
\item \emph{Failure correlation} is frequently disregarded. The assumption that failures are independent and that time between failures is exponentially distributed is not valid~\cite{schroeder2007disk}. 
\end{enumerate*}
\noindent

We propose to rethink disk arrays with the use of AE codes. 
The acronym RAID states for redundant arrays of independent\footnote{Patterson et al. used the word ``inexpensive'' but solutions became complex and expensive and ``independent'' was gradually adopted.} disks. 
Given that AE codes creates interdependencies between devices, it makes sense to drop the term ``independent" and replace it with ``interdependent''.
Note that this use case is valid for log-structured append-only storage systems.\\
 
\subsubsection{Entangled Mirror} 
In earlier work, we proposed two different array organizations based on simple entanglements~\cite{estradasimple}. 
Both organizations require equal numbers of data and parity drives; therefore, the array has the same space overhead as mirroring. 
For the sake of completeness, a quick recap is included in this subsection.

\noindent
\textbf{Full Partition.}
In this approach, a node represents a data drive and an edge represents a parity drive.
Blocks are written sequentially on the same drive type. 
The process does not spread content across drives. 
Most of the drives will remain idle and can be powered off as in a massive arrays of idle disks, MAID, configuration~\cite{colarelli2002massive}, which could result in significant energy savings. \\
\noindent
\textbf{Block-level Striping.}
This second approach distributes data over all available drives to improve performance.\\
\indent
In an entanglement chain, blocks that are located at the extremities have less redundancy. 
This problem has more impact on full partition where the content that is at extremity of the chain is equivalent to the stored data in one disk. 
At block-level striping, the amount of data is equivalent to the block size. 
We proposed open and closed chains, the second approach to address this problem. 
We showed that in full partitions both approaches provide better 5-year reliability than mirroring, reducing the probability of data loss by respectively 90 and 98 percent. 
\subsubsection{RAID-AE} 
Storing data in disk arrays is often held as an outdated solution. 
Given the scalability issues of RAID and the growing cloud market, some organizations  move data to the cloud.  
However, in-house storage is still used by many industries, e.g. finance market, government, and research institutions. 
On the other side, many cloud storage services are built on infrastructures that use RAID-like storage solutions. 
For example, Backblaze's backup service uses a RAID system built on top of Linux and RS codes~\cite{backblaze_reedsolomon}. 

There are three ways to increase the redundancy of data when configuring disk arrays: mirroring (discussed previously), parities and erasure codes. 
RS codes became a sort of de-facto industry standard for erasure coding, and particularly to archive data. 
Perhaps the main reason for their broadly acceptance is because, when the storage market became to grow, they were already an established solution for digital communications. 
In brief, RS codes are storage efficient, well-understood, and open-source libraries are available.
But the bandwidth and I/O cost during the repair process is only partially solved with optimal locally repairable codes~\cite{papailiopoulos2014locally}.

We briefly discuss the properties of a RAID built with our entangled-based solution.\\
\noindent
\textbf{Scalability.}
RAID-AE can make improvements for both horizontal and vertical scalability. 
In a RAID-AE array, it is possible to add more disks and change the storage capacity by replacing disks or creating volumes. 
Both actions may be done dynamically without interrupting the service and without encoding data again. \\
\noindent
\textbf{Never-ending Stripe.}
Entanglement codes change the classical notion of \emph{stripe} due to the possibility for writing on a never-ending stripe (a lattice has no limits), on the other side, the write penalty is $\alpha + 1$.
For clear examples, we use RAID5 but with more complex RAID organizations the situation remains the same. 
The way to compute parities is based on a fixed-width stripe, e.g. in a 6+1 disk RAID5, one parity unit is computed using 6 data units.  
When one disk is added to the array, the new array 7+1 disk RAID5 requires re-encoding parities, one parity unit is now computed using 7 data units.
Additionally, there is a large penalty when a disk fails because of the bandwidth and I/O overhead to repair each missing unit. 
Parity declustering~\cite{holland1992parity} reduces the rebuild time by distributing repairs across devices in a large array but this technique cannot reduce the overhead mentioned above. 
We argue that the scope for major improvements is limited by the stripe size, increasing the length of a stripe has a prohibitive effect in repairing data.\\
\noindent
\textbf{Degraded Reads.}
The stripe size impacts on degraded reads in RAID-like solutions.
In data centre environments where disks are distributed in different machines and racks, the performance of degraded reads matters.
RAID-AE provides many alternative paths to read data that is temporarily unavailable due to software updates, schedule restarts, etc. \\
\noindent
\textbf{Other Features.}
RAID-AE can be implemented to provide distributed repairs and load balance for read intensive workloads.   
One important difference in our model is that we do not assume failure independence and because the parameter $\alpha$ can change in future, the system can scale in fault tolerance. 

In sum, RAID-AE will permit different arrangements with trade-offs between capacity overhead, network bandwidth overhead, rebuild time and fault tolerance.

%% file: evaluation.tex

\section{Evaluation}
\label{evaluation}
We examine the design of AE codes to understand how the code settings impact on fault tolerance and write performance. 
Then, we compare AE codes with traditional redundancy schemes (RS codes and replication) in disaster recovery by measuring data loss and vulnerable data.
RS codes conceptualize the idea of an ``ideal code'' that has optimal characteristics and for that reason can be used as a baseline to compare with other codes~\cite{hafner2004performance}. 
In addition, RS codes are well-understood and many other codes are built on top of them. 
Given the radical differences of AE codes with other codes the comparison is not a trivial task. 
AE codes construct a scalable fabric of interdependent data that propagates redundancy, and this overlay can be mapped to the physical storage layer.
To the best of our knowledge, our notion of redundancy propagation has never been explored at this level in other codes.   
Oversimplification of our model would lead to unfair results and adding more codes in the comparison would create unnecessary complications. 
Instead, we design metrics that focus on the main characteristics presented in the previous sections to study them from different angles.  
To complete this study, we draw attention to the hidden role of maintenance and the risk of data loss in unreliable environments. 
We investigate how much redundant data are needed in unreliable environments.
Finally, we show evidence on the impact of data placements for different redundancy schemes.

\subsection{Code Parameters and Fault Tolerance}
\begin{figure}[t!]
\centering
\includegraphics[scale=0.31]{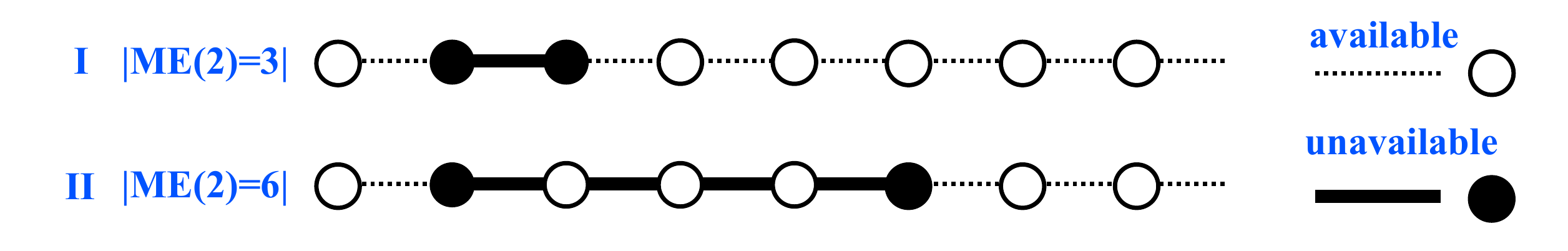}\\
\caption{\emph{Primitive forms.} Single entanglements cannot tolerate I) triple failures affecting two adjacent nodes and their incident shared edge and II) its extended form in which the nodes are not adjacent but all the connecting edges are unavailable.}
\label{fig:primitive_forms}
\end{figure}
\begin{figure}[t!]
\centering
\includegraphics[scale=0.31]{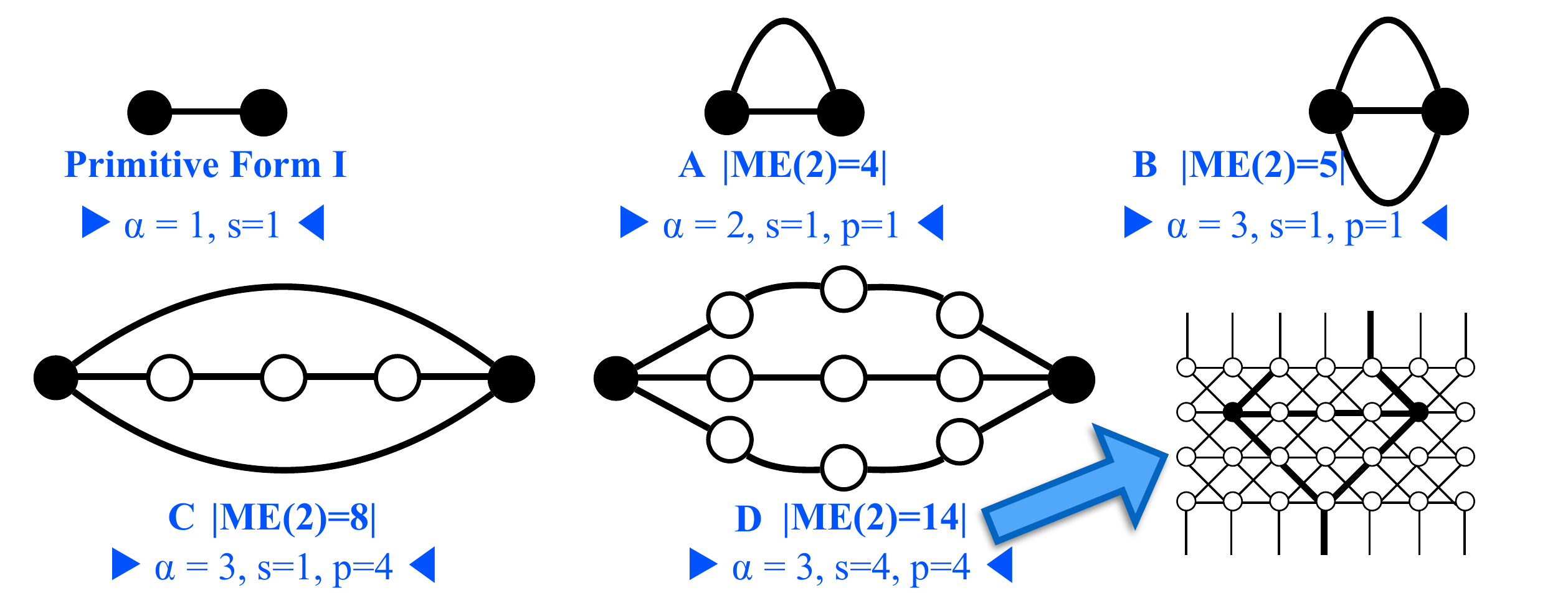}\\
\caption{\emph{Complex forms.} When $\alpha \geq 2$ primitive forms do not cause data loss only their combinations into complex forms as shown in patterns A-D.}   
\label{fig:complex_forms}
\end{figure}
\begin{figure}[t!]
\centering
\includegraphics[scale=0.38]{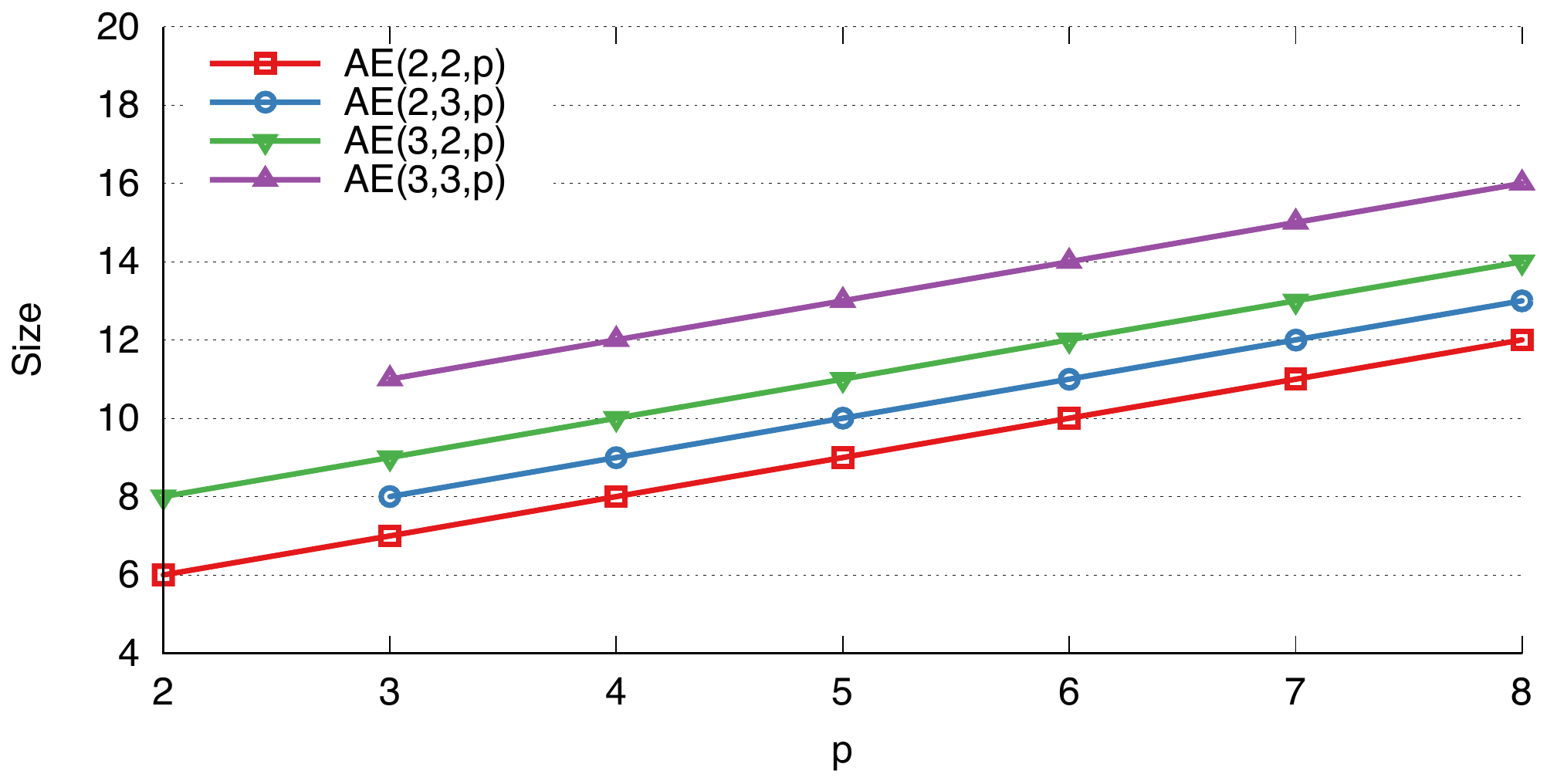}\\
\caption{\textbf{$|\textrm{ME}(2)|$} increases with larger s and p.}
\label{fig:sizeME_2}
\end{figure}

\begin{figure}[t!]
\centering
\includegraphics[scale=0.38]{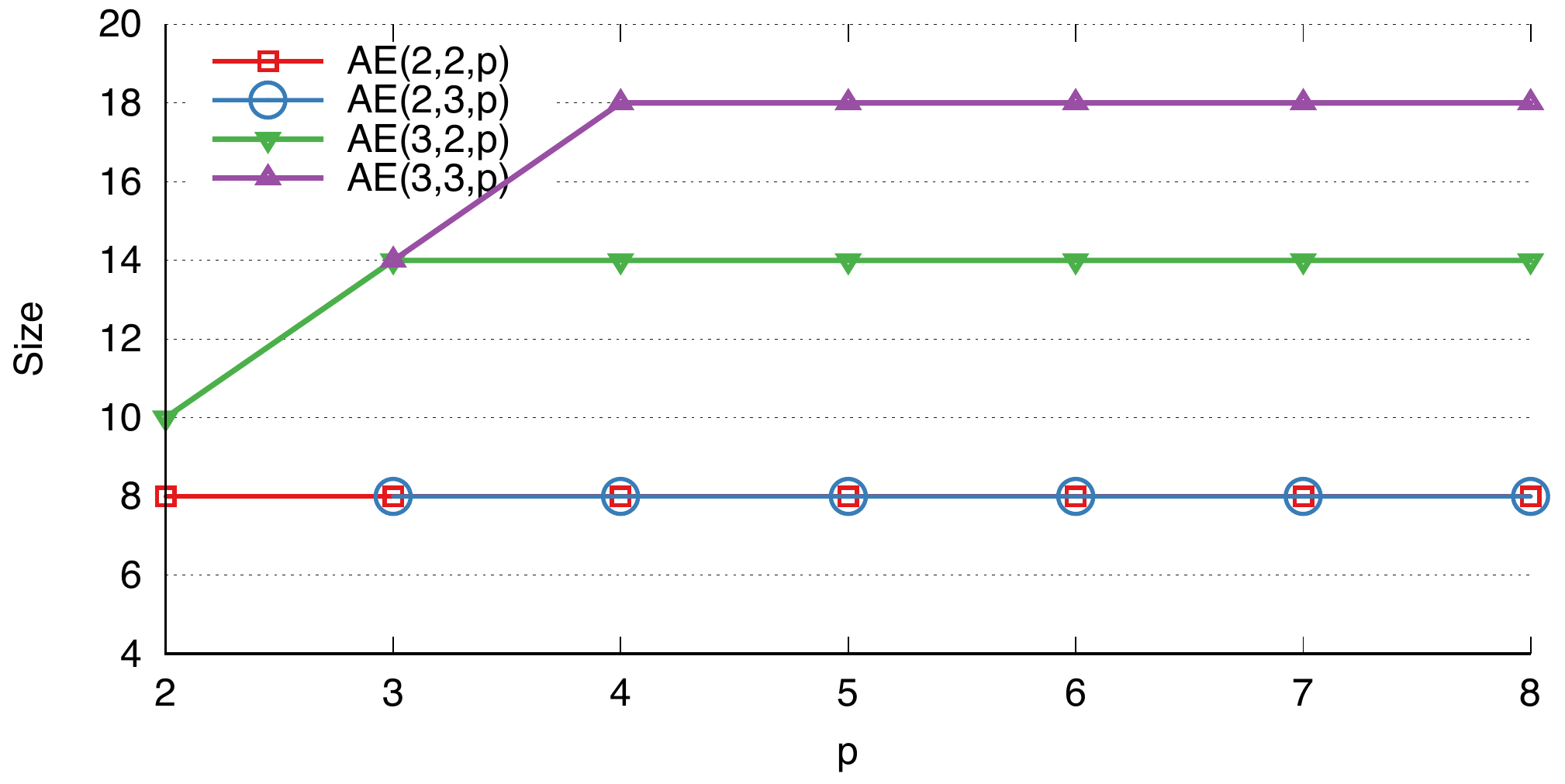}\\
\caption{\textbf{$|\textrm{ME}(4)|$} remains constant for $\alpha=2$, and increases with s for $\alpha=3$.}
\label{fig:sizeME_4}
\end{figure}

We have conducted a large study on the erasure patterns that a helical lattice cannot tolerate. 
The motivation was to understand how the code parameters impact on the fault tolerance. 
More specifically, we show visual evidence of how the parameters $s$ and $p$ increase fault tolerance without the need of additional storage or bandwidth. 

A minimal erasure (ME) is an irreducible pattern that causes the loss of data and parity blocks. 
These patterns are irreducible in the sense that the removal of any of their blocks enables the possibility to recover some of the erased blocks. 
The MEL~\cite{wylie2007determining} is the enumeration of all minimal erasures in irregular XOR-based flat codes. 
We use a variation of Wiley's study to gain more knowledge about the pattern size and its impact on data block loss.
Wiley focused on characterizing failure patterns  by their size without making the distinction between pattern size and actual data loss, or the impact that code parameters have in those failure patterns. 
Our goal is to characterize minimal erasure patterns by their size and amount of data loss.  
Our modification includes the notion that an erasure pattern of size $y$ blocks only has $x$ data blocks, with $y>x$. 
\emph{Ideally, we want patterns with $y \gg x$ since that means a high fault-tolerance, which decreases the probability that the decoder fails, but when it fails only a small fraction of blocks are data blocks}. 
For many patterns we can increase the ratio $y/x$ using the code parameters. 
An outstanding characteristic is that the parameters $s$ and $p$ permit increasing the ratio $y/x$ without generating more storage overhead or increasing the repair cost of single failures.
The increase in fault tolerance when tuning the parameters is evidenced by a cross-study comparison of the size of minimal erasure patterns for different code parameters. 
This study does not identify all erasure patterns for $\alpha$-entanglements. 
To minimize the burden of such task, we concentrate only on the most relevant patterns to determine which of them have lower and upper bounds for the redundancy propagation achieved by entanglement codes. 
Fig.~\ref{fig:primitive_forms} and~\ref{fig:complex_forms} present examples for ME(2). 
Our findings come from visual inspections and from verifications conducted with a tool\footnote{Available at request.} implemented in the Prolog language.   

\noindent
\textbf{Results.}
Fig.~\ref{fig:sizeME_2} and~\ref{fig:sizeME_4} show that $|\textrm{ME}(x)|$ is minimal when $s = p$, i.e. the code provides the minimum possible fault-tolerance for a given $\alpha$ and $s$. 
Although the code can provide more fault-tolerance when $p$ increases, the setting $s=p$ cannot be judge as good or bad per se. 
Trade-offs between performance and reliability should be considered to choose the best parameters.
Fig.~\ref{fig:sizeME_4} shows the impact of $\alpha$ on patterns ME($2^\alpha$). 
In this case, the parameters $s$ and $p$ have insignificant impact on fault tolerance. 
The reason behind this behaviour is that with $\alpha=2$ redundancy is propagated across elements that form a square pattern  (4 nodes and 4 edges, hence $|\textrm{ME}(4)|=8$) and these elements cannot be repaired if all are lost. 
With $\alpha=3$, $|ME(4)|$ becomes larger with $s$ but not with $p$, in this dimension, redundancy is propagated across a cube pattern, hence $|\textrm{ME}(8)|=20$ for AE(3,3,3), not shown for space reasons. 
  
\noindent
\textbf{Beyond $\alpha=3$.}
We are still investigating entanglement codes with $\alpha>3$.
We can safely speculate that the fault-tolerance would improve substantially.
We expect to find upper bounds for fault tolerance defined by the size of an $n$-hypercube.
This upper bound has impact on $|\textrm{ME}(x)|$ with $x=2^\alpha$, for example, for $x=16$ redundancy propagation would be enclosed in a tesseract (4-cube).
In addition, for larger $\alpha$ values we expect improvements in repair performance.  
However, it is not clear how to connect the extra helical strands.
In double- and triple-entanglements the RH and LH strands are defined along a diagonal of slope 1. 
One possible option is to add additional helical strands with a different slope. 

\subsection{Code Parameters and Write Performance}
\begin{figure}[t!]
\centering
\includegraphics[scale=0.50]{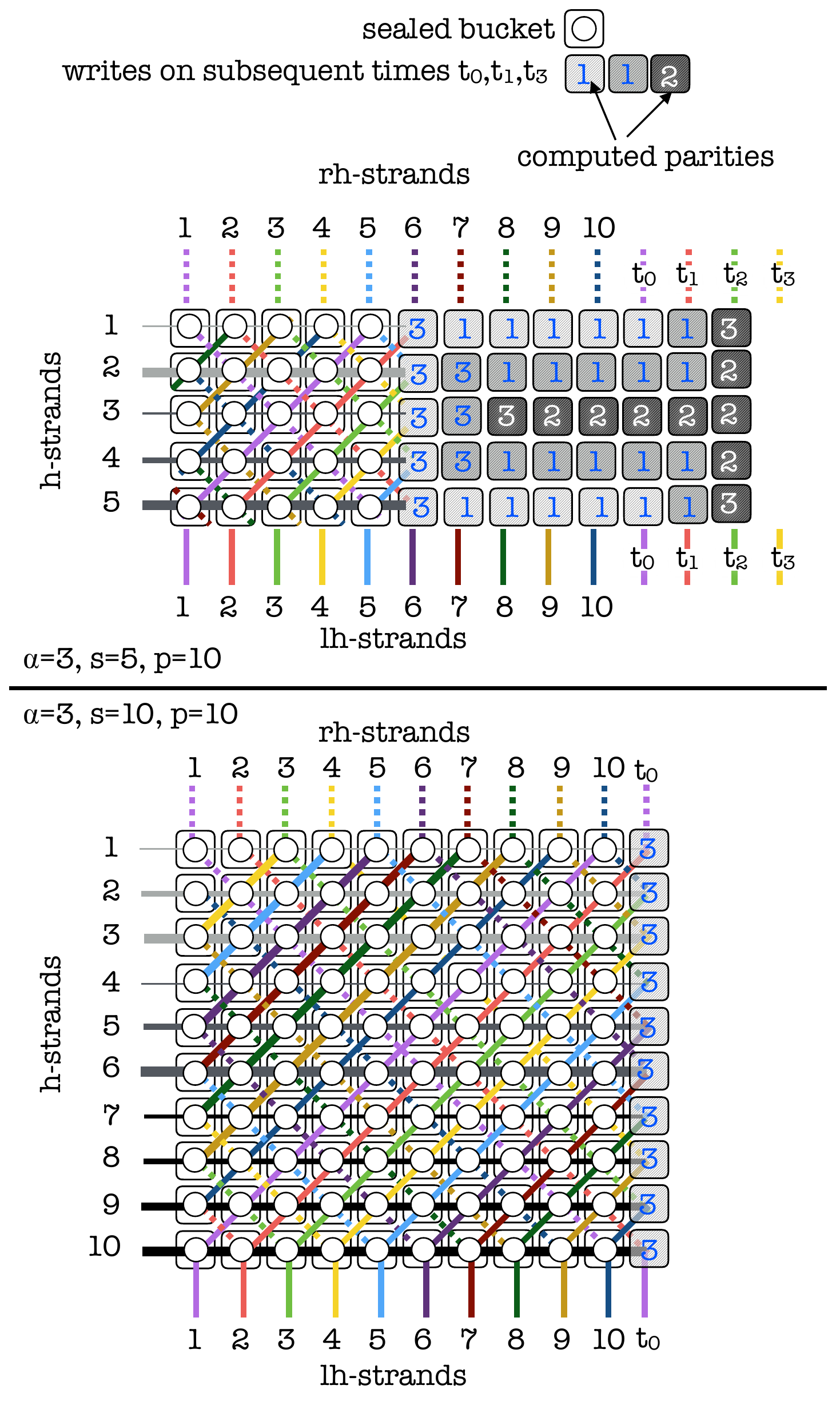}\\
\caption{Write performance for $p>s$ and $s=p$. Full-writes are optimized when $s=p$  since all the parities needed to complete triple-entanglements and seal the buckets are available in memory. When $p>s$, one option is to do full-writes for $s$ elements, a second option is to write buckets partially using the available parities.}
\label{fig:parallel_write}
\end{figure}
The values of parameters $s$ and $p$ impact on the number of data blocks that need to wait to be entangled.  
When $s=p$, this number is maximized and entanglements can be done in parallel operations, see Fig.~\ref{fig:parallel_write}. 
A \emph{sealed bucket} contains a data block and the $\alpha$ parities created by the entanglement process. 
A data block can be fully entangled when the $\alpha$ input parities needed in the process are already computed and kept in memory.
The memory requirement for full-writes is $O(N)$, where $N$ is the number of parities computed in the full-write.

\subsection{Simulations}
We run simulations to understand the scalable behavior of AE codes, assess them against catastrophic failures, compare them with $(k,m)$-codes and address the question of how much redundancy is needed to keep data safe in unreliable environments that have minimal (or zero) maintenance, i.e., repairs and efforts to restore redundancy after failures. 

\noindent
\textbf{Examples of Unreliable Environments.}
An unreliable environment could be a peer-to-peer (p2p) network where nodes join and leave frequently. 
Redundancy helps to protect data from real node departures and for maintaining high availability when hosts are offline. 
A relevant work on p2p storage showed a large-scale cooperative storage is limited by unreasonable cross-system bandwidth~\cite{blake2003high}.
Although this study from 2003 has some results that need update like current hardware trends and Internet connections, there are interesting insights from real world p2p networks. 
Nodes availability in a p2p network is very variable and maintenance swallows up most of the node's resources. 
Another relevant remark from the same work is that a system needs up to 120 copies to achieve high availability (6 nines) using replication, while erasure codes reach 6 nines with only 15 times the original storage needs.   
In archival storage systems, data durability is an endeavour that depends on the engineering aspects of the system but on the economics too~\cite{adams2011using,rosenthal2012economics}.
From the perspective of data centers, hard disks are unreliable components that contribute to high maintenance cost. 
To give a rough idea, the annual cost due to hardware repairs (mostly due to hard disks failures) for a data center with more than 100,000 servers was estimated to be over a million dollar by Microsoft's researchers in 2010~\cite{vishwanath2010characterizing}. 

\begin{table*}[!tb]
\setlength{\tabcolsep}{2pt}
\small
\caption{Redundancy schemes (examples for trivial replication cases not shown). AS: additional storage, SF: single failures}
\centering
\begin{tabular}{l | c | c| c| c c c c c c c}
\hline
\textbf{Cost} &\textbf{RS($k,m$)}&\textbf{AE($\alpha,s,p$)}&\textbf{$n$-way replication}& \textbf{RS(10,4)} & \textbf{RS(8,2)} & \textbf{RS(5,5)} & \textbf{RS(4,12)} & \textbf{AE(1,-,-)} & \textbf{AE(2,2,5)} & \textbf{AE(3,2,5)}\\\hline
\textbf{AS}  &$\frac{m}{k}\cdot 100\%$ & $\alpha \cdot 100\%$& $(n-1) \cdot 100\%$ & 40\% & 25\% & 100\% & 300\% & 100\% & 200\%& 300\%\\\hdashline
\textbf{SF}  & k & 2& 1 & 10 & 8 & 5 & 4 & 2& 2& 2\\
\hline
\end{tabular}
\label{tab:codes}
\end{table*} 
\begin{table}[!tb]
\setlength{\tabcolsep}{2pt}
\small
\caption{AE: d-block $d_{26}$ and p-blocks $p_{21,26},p_{26,31}$, $p_{22,26},p_{26,35}$, $p_{25,26},p_{26,32}$. Locations 3,12,47,56 are unavailable. Block $d_{26}$ is repaired via RH strand's p-blocks.}
\centering
\begin{tabular}{l c c c c c}
\hline
\textbf{i} & \textbf{j} & \textbf{Type/Strand} & \textbf{Location} & \textbf{Available}& \textbf{Repaired}\\\hline
26  & 26& d & 56 & FALSE& \textbf{TRUE} \\\hdashline
21  & 26& h & 3 & FALSE& TRUE \\\hdashline
26  & 31& h & 47 & FALSE& FALSE \\\hdashline
22  & 26& lh & 12 & FALSE& FALSE \\\hdashline
26  & 35& lh & 28 & TRUE& FALSE \\\hdashline
25  & 26& rh & 91 & \textbf{TRUE}& FALSE \\\hdashline
26  & 32& rh & 39 & \textbf{TRUE}& FALSE \\
\hline
\end{tabular}
\label{tab:sim_AE}
\end{table}

\noindent
\textbf{Selected Redundancy Schemes.}
We study 4 different settings for Reed-Solomon codes and 3 different settings for alpha entanglement codes, see Table~\ref{tab:codes}. 
In addition, we compare up to 4-way replication since $300\%$ is the maximum additional storage considered in this paper. 
Replication requires much more storage to offer high fault tolerance but it does not have overheads for single failures. 
This evaluation substantially improves previous evaluations ~\cite{galinanes2015ensuring} since we include many more code settings, millions of blocks and placements. 

\noindent
\textbf{Simulation Environment.}
We investigate what happens with data in environments that either accumulate plenty of failures before repairs take place or a large number of failures happen all at once. 
The metrics described in this section follows the criteria used in an on-going project intended to help in the comparison of codes~\cite{estrada2017research}. 
All experiments are done with 1 million data blocks and the corresponding number of encoded blocks for each code setting, e.g. RS(10,4) generates 400,000 encoded blocks while RS(8,2) generates 250,000 encoded blocks. 
That means that the number of stripes is different for each code setting, e.g. RS(8,2) creates 125,000 stripes, and  RS(5,5) creates 200,000 stripes.
Blocks are synthetically generated and stored in tables adapted for each code method.
Table~\ref{tab:sim_AE} shows a simplified version of the AE table that uses three columns to identify the block, one column to determine the location, one column to determine the availability of the block and one final column to specify if the block is repaired.

\noindent
\textbf{Block Placements.} 
Blocks are distributed in $n$ locations using random placements, i.e., each block is assigned a random number from 0 to $n-1$. 
We present the results of simulations for medium size ($n=100$) storage system
For example, the 1 million data blocks and 0.4 million encoded blocks created with RS(10,4) are distributed to 100 locations with a mean of 14,000 blocks per site and a standard deviation $\sigma=130.88$.
When locations are selected at random, it can happen that blocks from the same RS stripe are assigned to the same location. 
In our previous example, in a total of 100,000 stripes, only 38,429 had their 14 blocks distributed to different locations. 
The rest of the stripes were distributed in locations (stripes): 8 (5), 9 (39), 10 (475), 11 (3,746), 12 (17,076), 13 (40,230).
As expected, blocks had a fairly balanced distribution by increasing $n$, e.g., 91,167 stripes had their 14 blocks in different locations with $n=1,000$.
Spreading blocks evenly over nodes is a well-known scalability problem that affect load balance and system's reliability. 
While, at first sight, the distribution of RS encoded blocks is not even for a medium size system, we have run other simulations with a larger number of distinct locations and the comparisons remain close to the ones presented here. 
An important remark is that the random distribution of alpha entanglement encoded blocks in a medium size system affects the code performance in a worse way. 
For AE(3,2,5), the equivalent of 5-HEC, a lattice section of 80 elements (data and parities) need to be distributed in different failure domains.
The number was obtained from previous work ~\cite{galinanes2015ensuring}, where we assumed a round robin placement policy. 
This requirement assures that neighbour elements have more chances to be available in repairs that involved multiple blocks.
We think a round robin placement might be difficult to implement.
Therefore, this study answers what happens if we use random placements. 
It is expected that the subset of 80 blocks will not be fairly spread over all 100 locations, but does it affect the ability of the code to recover from disasters? \\
\noindent
\textbf{Disaster Recovery.}
The framework simulates disasters by changing the availability of a certain number of locations (10-50\%) and trying to repair the missing data blocks. 
When locations are unavailable, e.g., due to a faulty storage device, the repair process has to handle a very large number of single failures since a single location failure cause damage to multiple positions in the helical lattice.

\subsubsection{Metric: Data Loss}
 \begin{figure}[t!]
\centering
\includegraphics[width=0.9\columnwidth]{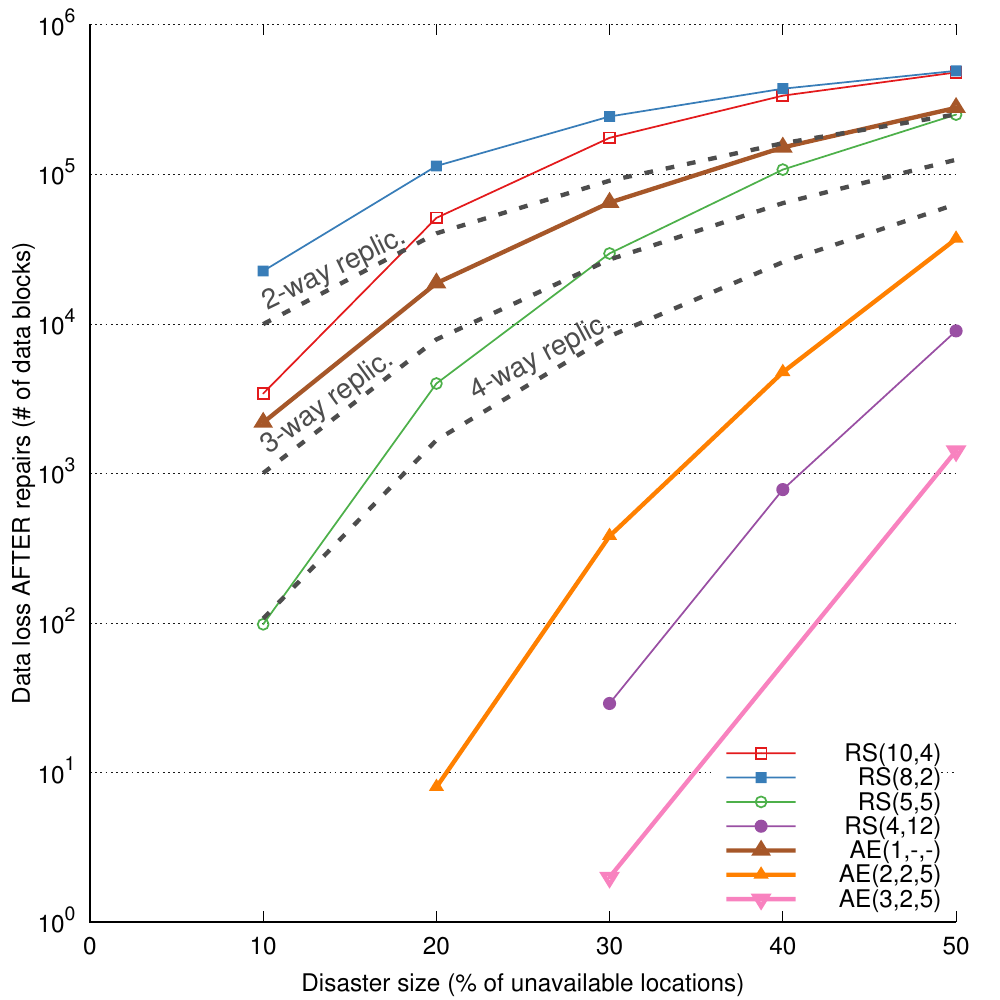}\\
\caption{Data blocks that the decoder failed to repair.}
\label{fig:data_loss}
\end{figure}
This metric only counts data blocks that satisfy the following conditions: a) its location is unavailable, b) the repair process was unsuccessful. 
Other available data blocks that belong to damaged stripes are not counted as lost. 
A RS stripe is damaged when more than $m$ blocks are unavailable. 
For example, a RS(8,2) stripe that has one encoded block and two data blocks in unavailable locations only contributes to the metric with two data blocks and the other six data blocks are considered available. 
In real-world implementations, it is probably that a RS damaged stripe cause more data loss than the one indicated with our metric. 
In AE codes, each block is encoded individually and data loss counts exactly how many data blocks are lost. 
Fig.~\ref{fig:data_loss} shows that AE(3,2,5) outperforms RS(4,12) even though both have the same storage overhead.
Data loss for \mbox{AE(1,-,-)} is one order more than RS(5,5), a code that uses the same storage overhead, but the gap between both curves decreases when the number of unavailable locations increases. 
In fact, the RS(5,5) curve shows an interesting behavior of RS codes. 
When the disaster size is relatively small ($10\%$), data loss of RS(5,5) is equivalent to 4-way replication. 
But the quality of RS declines with larger disasters, i.e., with $30\%$ of unavailable locations data loss of RS(5,5) is equivalent to 3-way replication and with $50\%$ of unavailable locations RS(5,5) causes the same data loss as 2-way replication.  
AE(2,2,5) excel at repairing blocks and the storage requirement is equivalent to using 3-way replication. 

\subsubsection{Metric: Vulnerable Data}
\begin{figure}[t!]
\centering
\includegraphics[width=0.9\columnwidth]{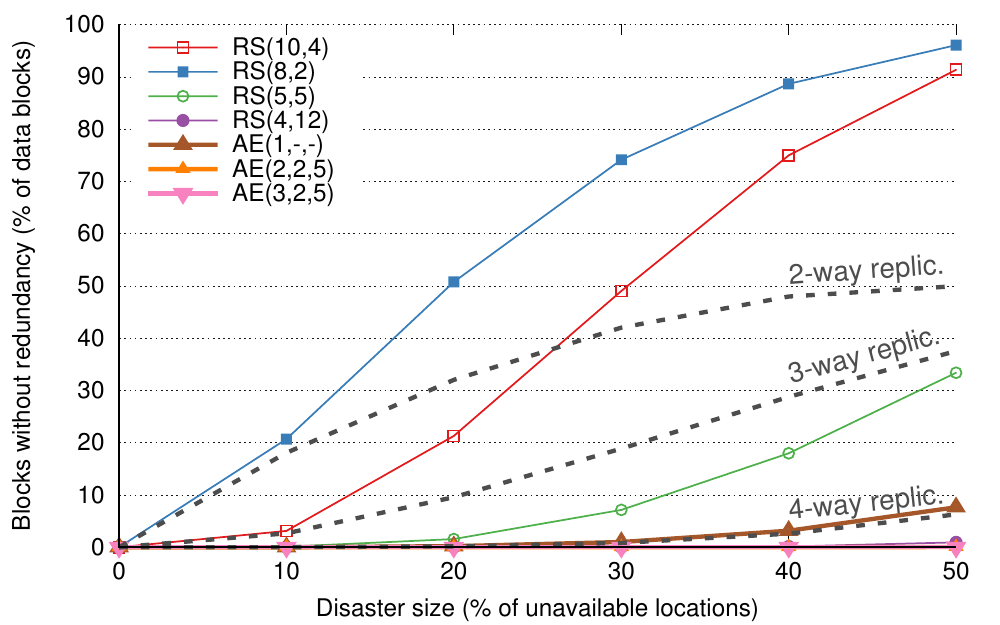}\\
\caption{Data blocks without redundancy.}
\label{fig:data_risk}
\end{figure}
This metric represents the total number of data blocks that are not protected by any other redundant block after the repair process finishes data recovering. 
It illustrates how the level of redundancy decays when only minimal maintenance operations are done in the system.
Minimal maintenance happens when the decoder repairs unavailable data blocks but make no attempts to repair unavailable parities. 
However, some parities are repaired if they are part of the same stripe of an unavailable data block.
A storage system can receive little maintenance under the following common scenarios: data block repairs are given priority, there is a lack of incentive to recover parities, missing parities may be difficult to detect when repairs are triggered by data block read failures.
Minimal maintenance can pose a threat to a big portion of the data in a storage system.
Fig.~\ref{fig:data_risk} shows the high percentage of data blocks that remain without redundancy after blocks repairs with RS codes. 
We observe that RS(5,5) performs worse than AE(1,-,-) and leaves more data without redundancy when failures affect more than 20\% of the locations. 
This result explains why in Fig.~\ref{fig:data_loss} the gap between the data loss curves for both codes becomes closer in large disaster scenarios.   
The RS(4,12) code is the only comparable to the high protection provided by AE codes.
Our results are somehow related with the impact of \emph{combinatorial effects} and \emph{peer availability effects} mentioned by other authors~\cite{lin2004erasure}.
These two effects explain the factors that benefit erasure codes and replication codes in p2p systems. 
Erasure codes take benefit of the combinatorial effect, however, they are more dependent of the availability of multiple locations. 
The authors added that when the peer availability is low, replication can be better than erasure codes. 
Alpha entanglements are less affected by the peer availability effect since only two blocks are used for repairs but they profit from the combinatorial effect as illustrated in Fig.~\ref{fig:novel_redundancy}.

\subsubsection{Metric: The Cost of Single Failures}
\begin{figure}[t!]
\centering
\includegraphics[scale=0.38]{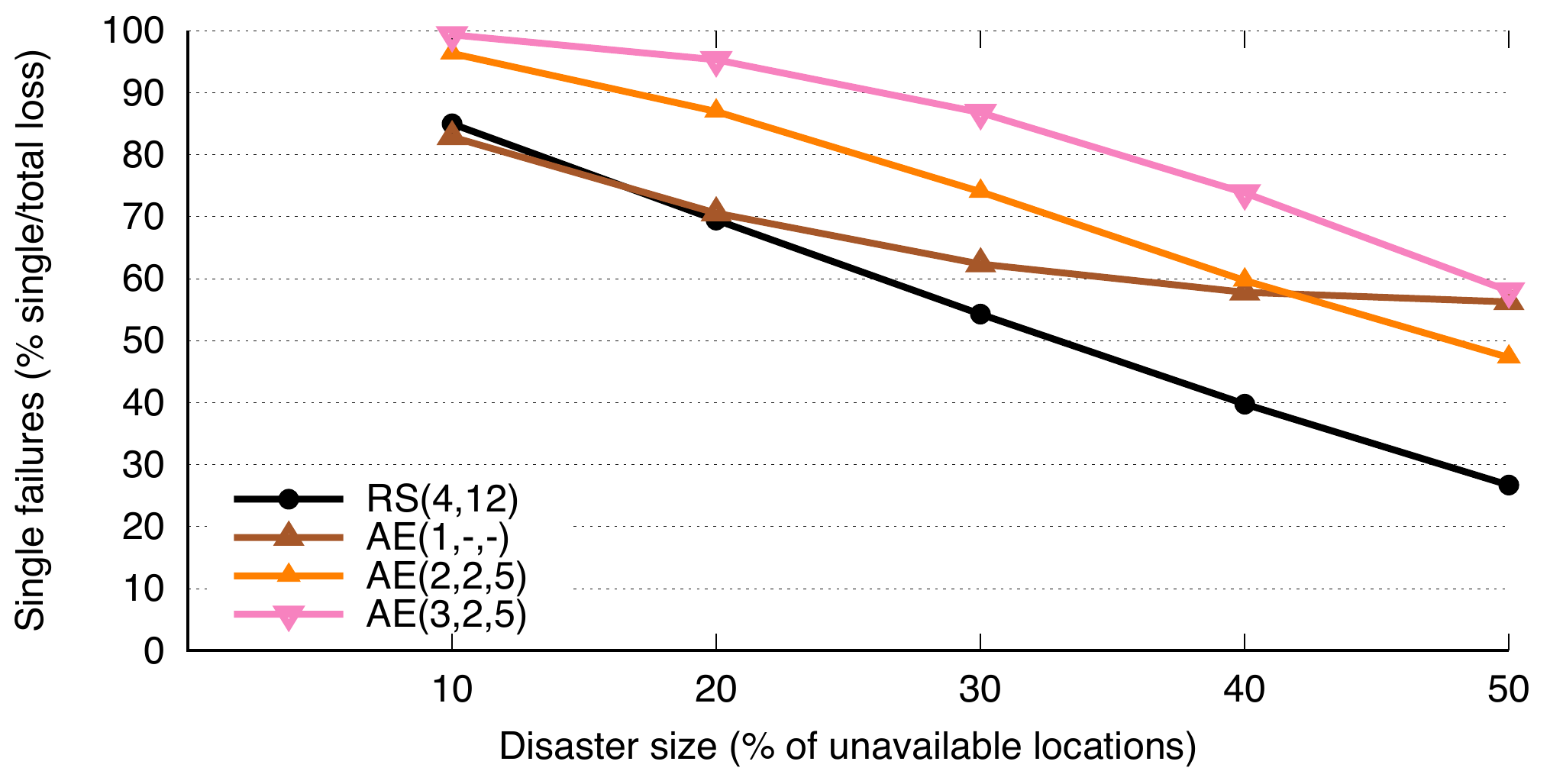}\\
\caption{What part of repairs are single failure repairs?}
\label{fig:single_failures}
\end{figure}

In \mbox{($k,m$)-codes}, repairing single failures (SF) is expensive. 
SF require $k$ times bandwidth overhead and I/O operations at $k$ locations. 
Alpha entanglement codes handle single failures with a radically different approach.
SF are repaired with a fixed ``k=2'' (two parity blocks) for any code setting.
However, repairing the full system may require several rounds.
Fig.~\ref{fig:single_failures} shows the percentage of data blocks that required SF repairs.  
For AE codes, we computed the proportion between all single failures solved at the first round and the total data loss repaired in the system.
For reference, we compute the same proportion (without considering rounds) for RS(4,12), which is the code with highest locality among our selected RS codes and superior than other locally repairable codes like the HDFS-Xorbas implementation~\cite{sathiamoorthy2013xoring}. 
For RS codes, the repair efficiency is very bad for small disasters. 
It improves for larger disasters because the number of single failures decreases.

\subsubsection{Metric: Code Performance}
At each round, our AE decoder computes 1 XOR between two available blocks for any data and parity blocks that is repaired. 
When data blocks cannot be repaired at the first round, the decode will do it at the second round if other required data or parity block becomes available. 
Fig.~\ref{fig:single_failures} shows that most data are repaired at the first round. 
The amount of blocks that is repaired per round decreases abruptly, normally the last round repairs only 1-2 blocks.
The number of rounds that were needed to repair all data blocks are summarized in Table~\ref{tab:code_performance}.
\begin{table}[t!]
\setlength{\tabcolsep}{2pt}
\small
\caption{AE codes: Number of repair rounds.}
\centering
\begin{tabular}{l c c c c c}
\hline
\textbf{Code} & \textbf{10\%} & \textbf{20\%} & \textbf{30\%} & \textbf{40\%}& \textbf{50\%}\\\hline
AE(1,-,-)  & 6& 7 & 9 & 10 &10\\\hdashline
AE(2,2,5)  &3& 6& 9 & 17 & 30 \\\hdashline
AE(3,2,5)  & 3& 4 & 7 & 10  &15\\
\hline
\end{tabular}
\label{tab:code_performance}
\end{table} 

%% file: related.tex

\section{Related Work}
\label{sec:related}
\noindent
\textbf{Data Entanglement in Censorship Resistant Systems.} 
The main spirit is the creation of a legal conundrum with the hope that a censor will not delete documents if a vast number of legal documents must be deleted to permanently delete an illegal document. 
If the system creates strong dependencies among the files, data loss can be recovered by exploiting the relationships between files.
The idea was explored in Tangler~\cite{waldman2001tangler}, Dagster~\cite{stubblefield2001dagster}, STEP-archival~\cite{mercier2015step} and Entangled Cloud~\cite{ateniese2016entangled}.
Aspnes et al. studied the theoretical aspects~\cite{aspnes2007towards}.

\noindent
\textbf{Data Entanglement to Reduce Storage.} 
Friendstore~\cite{Lillibridge:2003:CIB:1247340.1247343} is a cooperative backup system. Its redundancy scheme is called xor(1,2). It provides simultaneously redundancy for files owned by different users. It only encodes data when disk space is limited. The novelty was to trade-off bandwidth for storage. 

\noindent
\textbf{RAID.}
Disks arrays are organized in diverse ways: mirroring, RAID level 5~\cite{patterson1988case,chen1994raid}, RAID level 6~\cite{schwarz1992raid,burkhard1993disk}, two-dimensional RAID arrays~\cite{hellerstein1994coding, schwarz1994reliability}. 
Single-entanglements are close to SSpiral codes~\cite{amer2007outshining, amer2008increased}. 
Row-Diagonal Parity, also known as RAID-DP, protects data against double disk failures~\cite{corbett2004row}.
Researchers explored RAID at a petabyte scale~\cite{fan2009diskreduce}. 

\noindent
\textbf{Other Relevant Codes.} 
Parallel concatenated convolutional codes~\cite{benedetto1996design} and staircase codes~\cite{DBLP:journals/corr/abs-1201-4106} have some similarities with our work. 
Pyramid codes~\cite{huang2013pyramid} combine local and global parities to increase read efficiency during failures. 
RESAR uses a virtual layer to survive double-failures with low storage overhead and efficient updates~\cite{schwarz16-mascots}. 
LDPC codes~\cite{gallager1962low,luby1997practical} and regeneration codes~\cite{suh2009exact} have valued properties for efficient repair but they are not widely understood by the system community and  require complex parameterization~\cite{plank2003practical, jiekak2013regenerating}. 

%% file: conclusion.tex

\section{Conclusion}
Alpha entanglement codes are practical and flexible erasure codes designed to increase the reliability and integrity of a storage system, primarily for archiving data in unreliable environments. 
These mechanisms, based on the novel redundancy propagation, are not only very robust and sound, but they are also efficient and easily implementable in real systems. 
The encoder and decoder are lightweight---essentially based on exclusive-or operations---and offer promising trade-offs between security, resource usage and performance. 
We describe use cases for distributed and centralised storage systems.

The current research trend is to reduce to the maximum possible the storage overhead while improving the efficiency of repairs.  
While reducing storage costs satisfies industry needs, other problems are neglected. 
If we consider the potential savings on maintenance cost, AE codes make sense in a wide variety of scenarios whether they are decentralised or not.